\documentclass{icrc2009}

\usepackage{graphicx}   
\usepackage[caption=false]{caption}    
\usepackage[font=footnotesize]{subfig} 
\usepackage{fixltx2e}
\usepackage{url}

\newcommand{\shorttitle}[1]%
{\markboth{Proceedings of the 31\MakeLowercase{$^{st}$} ICRC, {\L}\'{o}d\'{z} 2009}{#1} }

\newcommand{\degr}{$^{\circ}$}

\begin{document}
\title{Search for Short Bursts of Gamma Rays Above 100~MeV from the Crab using VERITAS and SGARFACE}

\author{\IEEEauthorblockN{M. Schroedter\IEEEauthorrefmark{1} for the 
			  VERITAS Collaboration\IEEEauthorrefmark{2}}
                            \\
\IEEEauthorblockA{\IEEEauthorrefmark{1}Dept. of Physics and Astronomy, Iowa State University, Ames, IA, 50011, USA (schroedt@iastate.edu)}
\IEEEauthorblockA{\IEEEauthorrefmark{2}see R.A. Ong et al. (these proceedings) or http://veritas.sao.arizona.edu/conferences/authors?icrc2009 }
}

\shorttitle{M. Schroedter, Search for Bursts }
\maketitle

\begin{abstract}
The phenomenon of giant radio pulses (GRP) from the Crab Pulsar can be studied at gamma-ray energies using atmospheric-Cherenkov telescopes such as VERITAS and the SGARFACE experiment attached to the Whipple 10 m telescope. Although these instruments are generally used for very-high-energy gamma-ray astronomy above 100 GeV, they also provide substantial sensitivity to short bursts of photons above 100~MeV lasting up to 15~$\mu$s. Motivated by the theoretical predictions for short microsecond-scale GeV bursts as counterparts to GRPs~\cite{Lyutikov2007}, we report on a search for gamma-ray emission using simultaneous observations of the Crab Pulsar taken with VERITAS and the SGARFACE experiment.
\end{abstract}


\section{Introduction}
Short bursts of gamma rays above 100~MeV and with duration less than a few $\mu$s might be produced by pulsars~\cite{Lyutikov2007}, by explosions of primordial black holes~\cite{Hawking1974}, or via some as-yet undiscovered mechanism, for example during hard gamma-ray bursts. 

The Crab pulsar was discovered by occasional outbursts of giant radio pulses (GRP)~\cite{Staelin1968}. The period of the Crab pulsar is 33~ms and the pulse profile features two dominant pulses called the main and inter-pulse. GRPs have been detected only during those phases~\cite{Lundgren1995}. The intensity of GRPs is variable and can reach orders of magnitude over the average period-integrated flux. It is interesting to note that the duration of GRPs also shows extreme variability, lasting between 2~ns and about 1~$\mu$s above 5~GHz~\cite{Hankins2003}. Enhanced optical emission during GRPs has been reported by~\cite{Shearer2003}, but no such correlation has been found at higher energies~\cite{Lundgren1995}. It has been proposed that $\sim$30~GeV gamma rays may be produced by curvature radiation during periods of GRPs~\cite{Lyutikov2007}. 

Motivated by the extreme variability of GRPs and the prediction for 30~GeV gamma rays at a fluence level detectable by imaging atmospheric-Cherenkov telescopes (IACTs), we searched archival data from the Very Energetic Radiation Imaging Telescope Array System (VERITAS)~\cite{Holder2006} and the Short Gamma Ray Front Air Cherenkov Experiment (SGARFACE)~\cite{Lebohec2005} for burst-like events. The analysis and upper limits over a range of burst durations and photon energies are given below.

\section{Technique}
Bursts of nearly simultaneous gamma rays above 100~MeV can be detected by their secondary Cherenkov radiation in the atmosphere using ground-based optical telescopes. SGARFACE piggy-backs on the Whipple 10~m imaging atmospheric-Cherenkov telescope~\cite{Kildea2007} to search for such bursts. To eliminate cosmic-ray background events that mimic the expected image parameters of bursts, we search for correlations with VERITAS. VERITAS is one of the most sensitive arrays of atmospheric-Cherenkov telescopes and is located about 7~km from the SGARFACE experiment. Bursts of gamma rays would appear identical to both instruments, while cosmic rays would not be detected simultaneously and produe different image paramters.

The Whipple 10~m telescope is located at 31.6804\degr\ latitude, 110.8790\degr\ W longitude, and 2312~m a.s.l. The camera of the Whipple 10~m telescope consists of 379 close-packed PMTs covering a 2.4\degr\ field of view. The PMT signals for SGARFACE are split off the Whipple 10~m signal cables via passive couplers. To reduce the cost and complexity of the SGARFACE system, clusters of seven nearest-neighbor pixels are summed into 55 channels, each viewing $\sim$0.36\degr\ of the sky. This pixelation is sufficient to image the extended Cherenkov images produced by bursts of gamma rays. The channels are sampled by flash analog-to-digital (FADC) converters at an interval of 20~ns and with memory depth of 35.04 $\mu$s (1752 samples). The FADC input was designed with a time-constant of about 25 ns to preserve the accumulated charge between samples. SGARFACE achieves a high level of background rejection of cosmic rays and atmospheric phenomena through time-resolved imaging.

The SGARFACE trigger consists of two systems: a multi-time-scale discriminator (MTD) and a pattern-sensitive logic. The MTD integrates the signal on six time scales: 60 ns, 180 ns, 540 ns, 1620 ns, 4860 ns, and 14580 ns. The trigger threshold requirement is tested at three equally spaced time intervals during the trigger window, requiring pulses to be at least as long as the trigger window. The thresholds were set conservatively above the night-sky background and occasionally adjusted to reflect changes in the telescope throughput. The pattern-sensitive coincidence logic determines if a sufficient number of nearest neighbor channels have triggered the MTD. Due to the broad angular extent of images produced by bursts of gamma rays, all observations were taken with the trigger requiring 7 nearest-neighbor pixels above threshold. In turn, this requirement reduces triggers by the narrower gamma-ray and cosmic-ray initiated showers. Once an event triggers, the 55 FADC memories are read out after a 16.28 $\mu$s delay. 

The fluence sensitivity for a range of burst durations and photon energies was determined with Monte Carlo simulations%
\footnote{We use the KASCADE~\cite{Kertzman1994} particle air-shower simulation program, version 7.3, together with GrISU(tah) to simulate the air-Cherenkov production and telescope response. See http://www.physics.utah.edu/gammaray/GrISU/}%
. 

\begin{figure}[!t]
 \centering
 \includegraphics[width=2.5in]{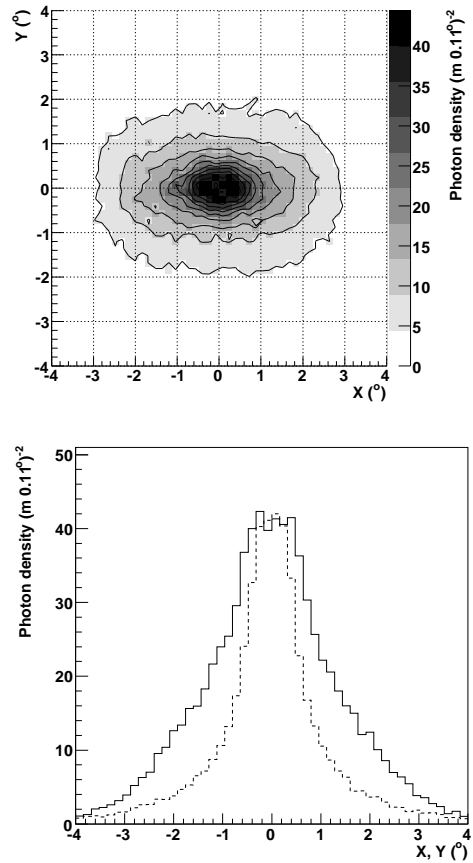}
 \caption{Angular distribution of Cherenkov light received at 1245 m altitude from a plane wave front of 1~GeV gamma rays at zenith with fluence of 2~$\gamma$/m$^2$. The size of the solid angle corresponds to the effective PMT cathode area over which 100\% efficiency would be achieved. The x-axis is perpendicular to the Earth's magnetic field. \emph{Lower figure:} The \emph{solid} line shows the Cherenkov photon distribution along the x-axis, the \emph{dashed} line along the y-axis. }
 \label{fig:veritas_distribution}
\end{figure}

\begin{figure}[!t]
 \centering
 \includegraphics[width=3in,clip]{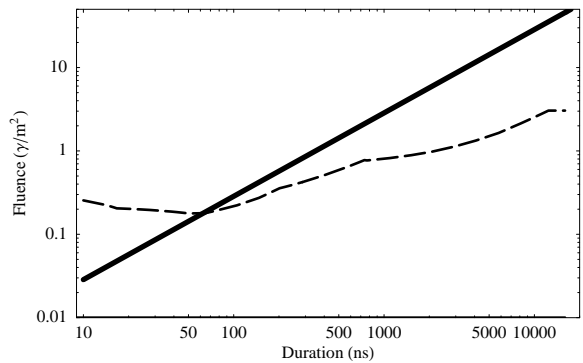}
 \caption{Minimum fluence required in bursts of 1~GeV gamma rays for detection with VERITAS (\emph{solid line)} and SGARFACE (\emph{dashed line}). The simulated bursts originate in the center of the field of view with the telescope pointing at zenith. The sensitivity varies with elevation and azimuth angle as shown in~\cite{Schroedter2009}.}
 \label{fig:veritas_sgarface}
\end{figure}

The sensitivity of VERITAS to bursts of gamma rays is set by the single photo-multiplier tube (PMT) trigger threshold of $\sim$5 photo-electrons%
\footnote{The four VERITAS telescopes have slightly different thresholds owing to the PMTs used in the camera. The threshold of the least sensitive telescope is $\sim$5.1 photoelectrons.}%
. Three nearest-neighbor PMTs, each viewing 0.15\degr\ of the sky, are required to exceed this threshold within 7~ns of each other.
The number of photoelectrons produced per Cherenkov photon is 0.14; this was determined by folding the simulated Cherenkov photon spectrum with the measured mirror reflectivity and the PMT quantum efficiency. The angular distribution of Cherenkov light was simulated for a burst of 1~GeV gamma rays with fluence of 2~$\gamma$/m$^2$ (Fig.~\ref{fig:veritas_distribution}). Combining this with the mirror area of $\sim$95~m$^2$ results in a minimum fluence required for the telescope to trigger of 0.02~$\gamma$/m$^2$ during 7~ns. The sensitivity to bursts of duration, $\tau$, decreases as $1/\tau$. Fig.~\ref{fig:veritas_sgarface} shows the minimum required fluence of a burst of 1~GeV gamma rays so that it is detected by VERITAS.

For SGARFACE, a detailed simulation was carried out that includes the measured single photo-electron probability distribution~\cite{Schroedter2009}. The minimum fluence sensitivity is shown in Fig.~\ref{fig:veritas_sgarface}.

\begin{figure}[!t]
 \centering
 \includegraphics[width=2.5in,clip]{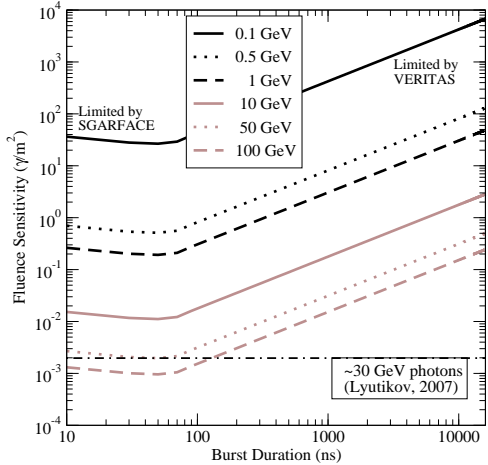}
 \caption{Sensitivity to detect bursts of gamma rays simultaneously with SGARFACE and VERITAS. The combined experiment has sufficient sensitivity to directly detect bursts of photons above 50~GeV at a fluence level predicted to arise from the Crab (\emph{dot-dashed line})~\cite{Lyutikov2007}. }
 \label{fig:sensitivity}
\end{figure}

The combination of VERITAS and SGARFACE results in a highly sensitive experiment over a range of gamma-ray energies, shown in Fig.~\ref{fig:sensitivity}. The space borne gamma-ray instrument \emph{Fermi} has good sensitivity to gamma rays above 0.03~GeV. We note that in order to detect a burst of gamma rays with \emph{Fermi}, a fluence of at least a few $\gamma$/m$^2$ would be required. The time scale for burst detection with \emph{Fermi} is limited to less than 1~$\mu$s for detection of multiple photons in a single event, and to greater than 26~$\mu$s because of instrument dead time.

\section{Data and Analysis}
Coincident observations of the Crab with SGARFACE and VERITAS have been carried out since October 2007 for a total of 6.3~hours taken under clear, dark skies and at telescope elevation above 60\degr.

\begin{figure}[!t]
 \centering
 \includegraphics[width=2.5in]{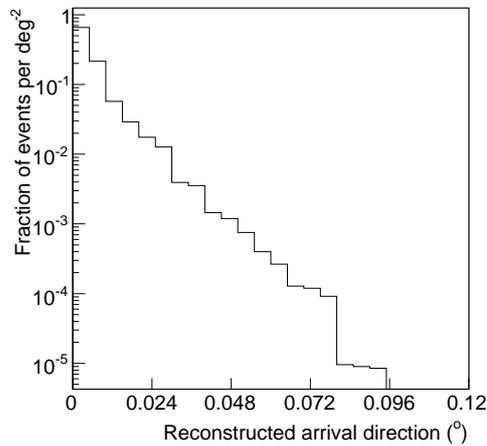}
 \caption{ Effective point spread function for bursts originating at the center of the field of view. The histogram shows the reconstructed burst direction for simulated bursts of gamma rays between 100~MeV and 10~GeV. The events pass the cuts of the SGARFACE analysis (see text).  }
 \label{fig:distance}
\end{figure}

SGARFACE data were analyzed as described in~\cite{Schroedter2009}. Briefly, the image shape and pixel timing are parameterized and events are then selected based on comparison to Monte Carlo simulations of bursts of gamma rays. The only difference from the cuts used in~\cite{Schroedter2009} is that the brightest pixel was not required to be below the FADC saturation level of $\sim$230 digital counts. A 98\% rejection efficiency of background from cosmic rays and atmospheric phenomena is achieved by this analysis.  Fig.~\ref{fig:distance} shows the SGARFACE point spread function for bursts of gamma rays between 100~MeV and 10~GeV and with duration between 10~ns and 15~$\mu$s. All events should fall within 0.1\degr\ from the true source location. Including the systematic telescope pointing error of 0.085\degr, the reconstructed burst direction is required to lie within 0.2\degr of the Crab. 

\begin{figure}[!t]
 \centering
 \includegraphics[width=2.5in]{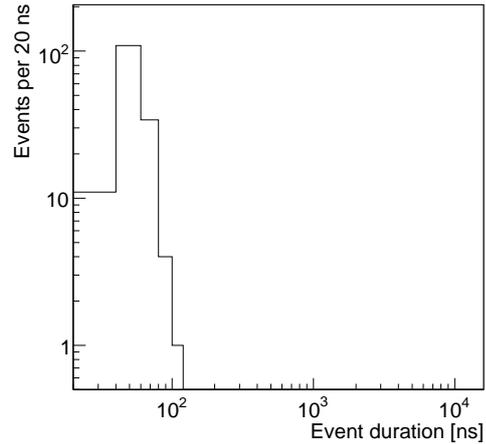}
 \caption{ Histogram of the duration of events measured by SGARFACE. These events have image and timing characteristics that are consistent with a burst of gamma rays coming from the Crab. The quantization of the duration in steps of 20~ns is the result of the analysis method.  }
 \label{fig:sgarface_events}
\end{figure}

A histogram of the event durations measured by SGARFACE is shown in Fig.~\ref{fig:sgarface_events}. The 159 events seen in SGARFACE are then searched for coincident events recorded by VERITAS. A 1~$\mu$s coincidence window, w, was used in searching for simultaneous events%
\footnote{The GPS time stamp reported by VERITAS for an event is averaged over all triggering telescopes. Since a burst of gamma rays would trigger all four telescopes, the event time stamps for nearly all events are accurate to better than 0.5~$\mu$s. Occasional glitches of $\pm$3~$\mu$s are known to occur in the GPS clocks of individual telescopes. }%
.\\


\begin{figure}[!t]
 \centering
 \includegraphics[width=2.5in,clip]{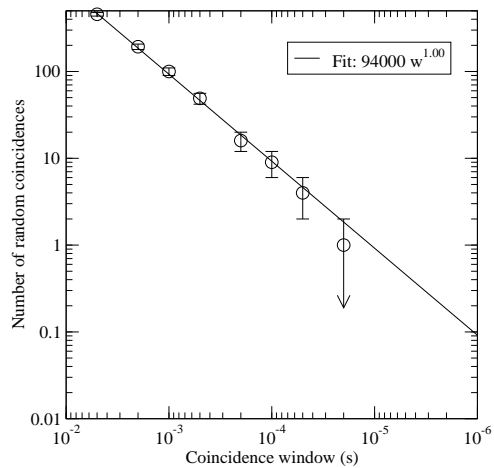}
 \caption{Number of random background events versus coincidence window, w. The number of random events was measured by offsetting the SGARFACE and VERITAS data sets by 1~s with respect to each other. The \emph{solid line} is a power-law fit to the data and shows the expected number of random coincidences for a 1~$\mu$s coincidence window to be 0.1. }
 \label{fig:random_coincidences}
\end{figure}

No simultaneous events were found between SGARFACE and VERITAS. The expected number of random coincidences for the data sets is 0.1, determined by offsetting the data sets from each other by 1~second and searching for coincidences at successively smaller coincidence windows as shown in Fig.~\ref{fig:random_coincidences}. With 0 measured bursts of gamma rays during 6.3 hours, the 99\% upper limit on the true rate of bursts is 0.73~hour$^{-1}$, assuming a Poisson distribution. This upper limit is conservative beyond 100~ns because SGARFACE is more sensitive than VERITAS, and no events longer than 100~ns were identified by SGARFACE.

\section{Conclusion}
Individual bursts of gamma rays from the Crab above 100~MeV can be detected with SGARFACE and VERITAS. The upper limits presented here on the rate of such bursts are the most stringent upper limits above 1~GeV and with duration shorter than 15~$\mu$s, see Fig.~\ref{fig:sensitivity}. The search was motivated by a prediction for $\sim$30~GeV photons from curvature radiation produced during GRPs~\cite{Lyutikov2007}. Though the predicted fluence is a factor of $\sim$2 below that achieved with this search, considerable uncertainty is involved in modeling the environment of the Crab pulsar. Pending the future operation of the Whipple 10~m telescope, additional observations may provide the opportunity for more detailed studies of gamma-ray emission concurrent with giant pulses.

\section{Acknowledgments}
This research is supported by grants from the US Department of Energy, the US National Science Foundation, and the Smithsonian Institution, by NSERC in Canada, by Science Foundation Ireland, and by STFC in the UK. We acknowledge the excellent work of the technical support staff at the FLWO and the collaborating institutions in the construction and operation of the instrument.


\begin{thebibliography}{99}

\bibitem{Lyutikov2007} M. Lyutikov, MNRAS, 381, 1190 (2007).
\bibitem{Hawking1974} S.W. Hawking, Nature, 248, 30 (1974).
\bibitem{Staelin1968} D. Staelin and E.C. Reifenstein, Science, 162, 1481 (1968).
\bibitem{Lundgren1995} S.C. Lundgren et al., ApJ, 453, 433 (1995).
\bibitem{Hankins2003} T.H. Hankins et al., Nature, 422, 141 (2003).
\bibitem{Shearer2003} A. Shearer et al., Science, 301, 493 (2003).
\bibitem{Holder2006} J. Holder et al., APh, 25, 31 (2006).
\bibitem{Lebohec2005} S. LeBohec, F. Krennrich, G. Sleege, APh, 23, 235 (2005).
\bibitem{Kildea2007} J. Kildea, APh, 28, 182 (2007).
\bibitem{Kertzman1994} M.P. Kertzman and G.H. Sembroski, NIM A, 343, 629 (1994).
\bibitem{Schroedter2009} M. Schroedter et al., APh, 31, 102 (2009).

\end{thebibliography}
\end{document}